\documentclass[aps,prl,twocolumn,groupedaddress,floatfix]{revtex4}
\usepackage[latin1]{inputenc}  
\usepackage{graphicx}

\newcommand\eps{\epsilon}

\newcommand\be{\begin{equation}}
\newcommand\nd{\end{equation}}
\newcommand\bed{\begin{displaymath}}
\newcommand\ndd{\end{displaymath}}

\newcommand\ba{\begin{array}}
\newcommand\ea{\end{array}}
\newcommand\bea{\begin{eqnarray}}
\newcommand\nda{\end{eqnarray}}

\renewcommand\Re{{\rm Re}\,}
\newcommand\We{{\rm We}\,}

\begin{document}


\title{A numerical cough machine}


\author{Cesar Pairetti}
\affiliation{Sorbonne Universit\'e and CNRS, Institut Jean Le Rond d'Alembert, UMR 7190, Paris, France}
\affiliation{Centro Internacional Mec\'anica Computacional (CONICET - UNL), Santa Fe, Argentina}
\affiliation{Facultad de Ciencias Exactas, Ingeniería y Agrimensura (UNR), Rosario, Argentina}

\author{Rapha\"el Villiers}
\affiliation{Sorbonne Universit\'e and CNRS, Institut Jean Le Rond d'Alembert, UMR 7190, Paris, France}

\author{St\'ephane Zaleski}
\affiliation{Sorbonne Universit\'e and CNRS, Institut Jean Le Rond d'Alembert, UMR 7190, Paris, France}
\affiliation{Institut Universitaire de France, IUF, Institut Jean Le Rond d'Alembert, UMR 7190, Paris, France}
\email[]{stephane.zaleski@sorbonne-universite.fr}


\date{\today}

\begin{abstract}
  We introduce a simplified model of physiological coughing or sneezing, in the form of a thin liquid layer subject to a rapid (30 m/s) air stream. The setup is simulated using the Volume-Of-Fluid method with octree mesh adaptation, the latter allowing grid sizes small enough to capture the Kolmogorov length scale. The results confirm the trend to an intermediate distribution between a Log-Normal and a Pareto distribution  $P(d) \propto d^{-3.3}$ for the distribution of droplet sizes in agreement with a previous re-analysis of experimental results by one of the authors. The mechanism of atomisation does not differ qualitatively from the multiphase mixing layer experiments and simulations. No mechanism for a bimodal distribution, also sometimes observed, is evidenced in these simulations. 
\end{abstract}


\maketitle

\label{s1}
Coughing and sneezing are two processes by which a large number of droplets of muco-salivary fluid are exhaled and subsequently travel large distances in the environment \cite{bourouiba2014violent,bourouiba2020turbulent}. These phenomena have acquired an acute interest in the context of the so-called aerosol transmission route of the Covid-19 pandemic \cite{bala2020}, but have been studied for near a century in the context of respiratory diseases in general \cite{bourouiba2020fluid}. 
Early investigations by Duguid  \cite{Duguid:1946dw} and Loudon \& Roberts \cite{loudon1967relation} of the number and size of droplets emitted in coughing and sneezing events have yielded rich data, incorporating very large numbers of droplets. A comprehensive analysis of the characteristics of these droplets, both in size and velocity, would be of immense interest, as it is a prerequisite for the modeling of the droplet cloud propagating further downstream from the mouth.
Rapid photographic imaging \cite{scharfman2016visualization} has revealed features similar to those observed in other {\em atomization} processes, including thin liquid sheets, ligaments and droplets. In this context, the statistical distribution of droplet sizes is of particular interest. Although the log-normal distribution has been frequently mentionned in connection with exhalations \cite{Nicas:2017eo,wells1955airborne} as well as other atomizing flows \cite{ling17} many other distributions $N(d)$ of the diameter $d$ have been put forward such as compound
gamma distributions \cite{Villermaux:2011ffa} and many others. A re-analysis of the data of Duguid and Loudon \& Roberts for sneezes has however revealed \cite{bala2020} a $N(d) \sim d^{-2}$ scaling over an impressive three orders of magnitude.
This scaling allows determining the proportion of millimeter-sized droplets that travel short distances and the proportion of much smaller droplets that can be incorporated in a turbulent puff or particle-laden jet and travel long distances as discussed in  ref. \cite{bourouiba2014violent,bala2020}. The fraction of the exhaled muco-salivary fluid in each class of droplet sizes may also be determined in this way and the probability of having  viral loads in each class can be inferred under adequate hypotheses, such as a homogeneous distribution of the virus in volume or surface.

In order to better understand the fluid mechanics of exhalations, King, Brock \& Lundell \cite{King:1985cn} have designed a physical model of violent exhalation that may be nicknamed a ``cough machine''. Air is flowing at high speed (from 10 to 30 m/s) in a rectangular-section duct, with a flow rate analog of the observed human cough. A thin layer of muco-salivary fluid is deposited at the bottom of the duct. While King, Brock \& Lundell use the cough machine to observe non-atomizing waves on the thin layer, it can also be used to simulate droplet formation at higher speeds and/or lower velocities. The device would then achieve atomization through a process similar to that of shear flow atomization of planar sheets \cite{gorokhovski2008modeling,Villermaux:2011ffa,agbaglah2017numerical,ling17,zandian2019length,ling2019two}. (See also the recent review of numerical approaches in \cite{juric}.) However there are important differences since the liquid layer is initially at rest and the airflow is impulsive. These differences could lead to different distributions of droplet sizes and velocities, which are the focus of the current investigation.  


We model the muco-salivary fluid and the surrounding gas as a Newtonian fluid (The muco salivary fluid is
non-newtonian but these effects only kick-in at very small scales). Thus the flow is described by the Navier-Stokes equations, which we solve by a simple finite volume discretization in the one-fluid numerical approach to two-phase flow\cite{tryggvason11}, using a Volume-Of-Fluid method for tracking fluid interfaces\cite{tryggvason11}, an octree grid and the Bell-Collela-Glaz advection scheme on staggered grids as described in \cite{Popinet03}, a Crank-Nicholson method for viscosity \cite{lagree2011granular}, and a height-function method for curvature and surface tension \cite{Popinet09}. The octree grid is refined using a wavelet estimate of the local error, as described in the self-documented code Basilisk (\url{http://basilisk.fr}). We use thirteen levels of refinement at the maximum, which results in the equivalent of $2^{39} \simeq 10^{13}$ or sixteen billion grid cells. However, thanks to the octree nature of the simulation, only 10-100 million grid cells are used in practice. 
\begin{figure*}[t]
\begin{center}
\includegraphics[width=0.8\textwidth]{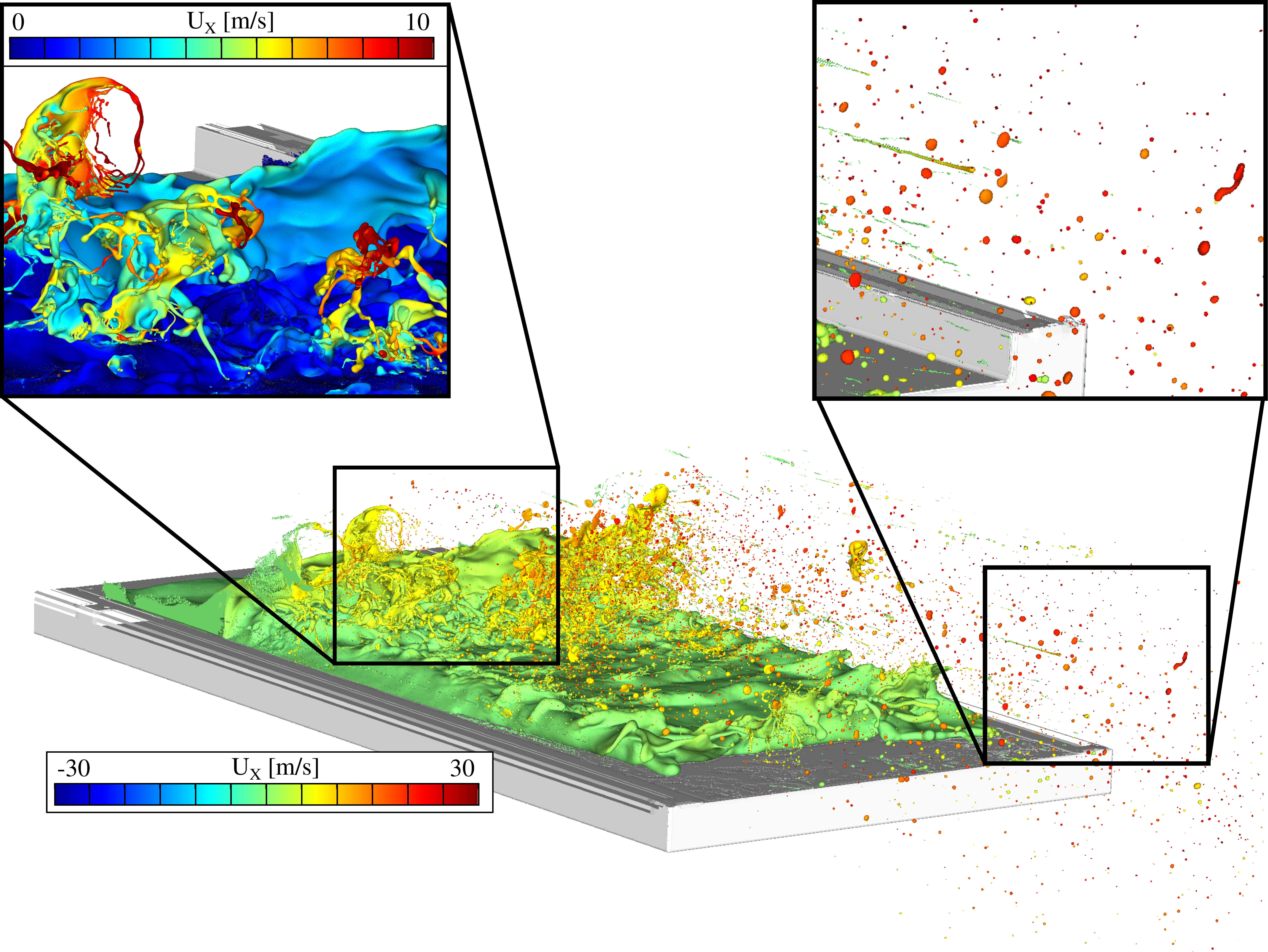}
\end{center}
\caption{A view of the air-liquid interface at t=9 ms. The interface is colored by the
  velocity of the droplets, with green the smallest velocity and dark red the largest.
\label{fig:droplets}}
\end{figure*}
The physical domain subject to the simulation is a cube of dimensions $L^3$ with flow inlet at $x=0$ and flow outlet at $x=L$. The flow is channeled through a tube of length $\ell_x$, and rectangular cross section with height $\ell_y$ and width $\ell_z$, with walls or plates of thickness $e$ centered on the planes $x=L/2 - \ell_z/2$ (bottom plate), $y=\pm \ell_y/2$ (lateral plate). The bottom plate is covered with a liquid phase of thickness $h$, density $\rho_l$ and viscosity $\mu_l$. The gas phase, modelling the exhaled air, has density $\rho_a$, and viscosity $\mu_a$. It enters the flow from the left wall at $x=0$ with uniform and constant velocity $U$ through an inflow boundary condition in the region $L/2 -\ell_y / 2 < y < L/2 +\ell_y / 2$ and $L/2 - \ell_z/2 + h < z < L + \ell_z/2$. This condition ensures the inflow of air blows just above the initial position of the liquid layer. The surface tension of the liquid is noted $\sigma$. Gravity is neglected 
as it is small compared to inertial effects $g \ell_z/U^2 \ll 1$.
\begin{table}
  \begin{tabular}{cccccccccccc}
    \hline \hline 
    $L$ & $\ell_x$ & $\ell_y$ & $\ell_z$ & $e$ & $h$ & $\rho_l$ & $\mu_l$ & $\rho_a$ & $\mu_a$ & $U$    & $\sigma$ \\ \hline
    0.15 & 0.05    & 0.01     &  0.02    & 0.002 & 0.001 & 1000 & 0.005 &  1.2 & $2\,10^{-5}$ & 30 & 0.03 \\
   \hline
   \hline \label{tabel}
  \end{tabular}
  \caption{Dimensional values of the fluid and geometrical parameters}
\end{table}
\mbox{}
The main dimensionless parameters are the Reynolds number of the air based on the channel height $\Re_a=\rho_a U \ell_z / \mu_a =36,000$, the Weber  number of the air based on the channel height $\We_a=\rho_a U^2 \ell_z / \sigma=720$, and the Reynolds number of the liquid based on the height of the liquid
 $\Re_l=\rho_l U h / \mu_l=30,000$. The values of these parameters are quite high which makes it surprising that the simulations, performed with a Navier--Stokes code, are still classified as  {\em Direct Numerical Simulations} (DNS). 
Indeed the assessment of the DNS character of the simulation may be inferred from the value of the dissipation $\epsilon$ in the similar setup of refs. \cite{ling17,ling2019two}. There the kinetic energy dissipation per unit volume $\epsilon$ \cite{pope2001turbulent} was measured and it was observed that the maximum value of $\eps/(\rho_a U^3/\ell_z)$ was about 0.01 which yields
an estimate of $\eps$. The Kolmogorov length scale is
then
$
\eta = (\rho_a \nu_a^3/\eps)^{1/4}
$
With the value of $\epsilon$ estimated above, we have $\eta_1 \simeq 24.2$ microns while the size
of the smallest grid cell with the thirteen levels of refinement in the simulation reported here
is $\Delta = 2^{-13}\, L \simeq 18.3$ microns. According to the DNS resolution criterion given by Pope \cite{pope2001turbulent}, the smallest turbulent scales will be well resolved if $\Delta/\eta \le 2.1$
while we have $\Delta/\eta_1  \simeq 0.75$
Even if one uses the more conservative estimate $\eta_2 = \ell_z \Re_a^{-3/4}$ as in \cite{juric} one gets
 $\Delta/\eta_2 \simeq 2.4$.
The perhaps surprising result that the simulations may be qualified as DNS can be explained by the fact that while the less extreme simulations of
 \cite{ling17,ling2019two} were limited to $\ell_z/\Delta = 256$ grid points in the gas jet thickness, our simulations using octree refinement go up to the equivalent of  $\ell_z/\Delta =  1092$. 

%
Simulations are initialized with zero velocity in the air and liquid although the incompressibility condition results in a non-zero velocity field everywhere in the gas immediately after time zero. The simulation is continued for a total time T=9 ms. The liquid surface is quickly significantly perturbed with waves present over the entire length $\ell_x$ of the tube, with a much larger wave near the inlet and some secondary waves near the outlet (Fig. \ref{fig:droplets}). The wave stretches into a thin liquid layer that fills with air under the combined effect of pressure and vorticity. Eventually, two mechanisms lead to the formation of droplets: the formation of fingers or ligaments at the end of the sheets and the puncturing of the sheets. The latter causes the formation of holes in the sheets and the subsequent expansion of those holes, leading to the formation of ligaments that eventually pinch and break into droplets. This process has been described in other experimental \cite{scharfman2016visualization} and numerical \cite{ling17} investigations. In the current ``closed channel'' configuration similar mechanisms of droplet formation are observed.
It is seen on Fig. \ref{fig:droplets} that the small dark red droplets near the channel exit have
moved quickly since the drag law
scaling for a sphere $F_d \sim \rho_a d^2 U^2/4$ (where $d$ is the droplet radius) implies that the
small droplets accelerate much faster than the large ones. 

\begin{figure}[!h]
\begin{center}
\includegraphics[width=0.35\textwidth]{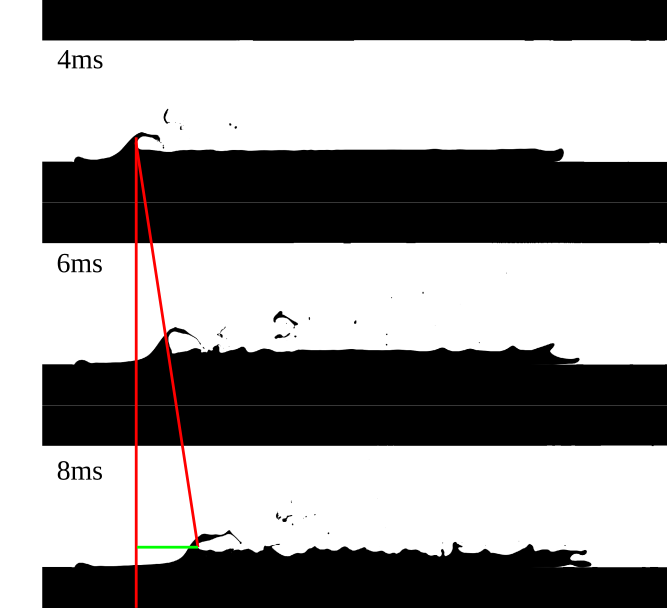}
\end{center}
\caption{In a cross section of the channel, the liquid phase and the solid are shown in black and the gas phase in white. Three different snapshots at regularly spaced time intervals are shown. The oblique line connects the positions, thus giving a graphical display of the wave velocity $U_D$, agreing with the Dimotakis velocity discussed in the text. }
\label{fig:wave}
\end{figure}
A clear phenomenon that couples with droplet formation is growth of a large sheet or wave near the inlet and its progression downstream. The velocity of such waves is typically the Dimotakis velocity
\cite{dimotakis86} and it is expressed as
$U_D = U{\sqrt{\rho_a} }/{(\sqrt{\rho_a} + \sqrt{\rho_l})}$.
A similar kind of solitary wave progressing at the Dimotakis velocity has been observed in the simpler setting of an infinite vortex sheet between air and liquid \cite{hoepffner11}. The liquid surface is excited by a local perturbation of the flat vortex sheet, and in this case the inhomogeneity of the flow at the entrance plays the role of the localised perturbation, while less localised waves are seen further downstream. In our case the Dimotakis velocity is $U_D \sim 1.004$ m/s which may be compared to a rough measurement from the simulations (Fig. \ref{fig:wave}) of $U_{D,\rm{num}} \simeq 1.25$ m/s. 
The identification of droplets or connected fluid components, and the computation of their volumes $V_d$ allows to define an equivalent droplet diameter $d = (6 V_d/\pi)^{1/3}$.
\begin{figure}[!h]
\begin{center}
\includegraphics[width=0.4\textwidth]{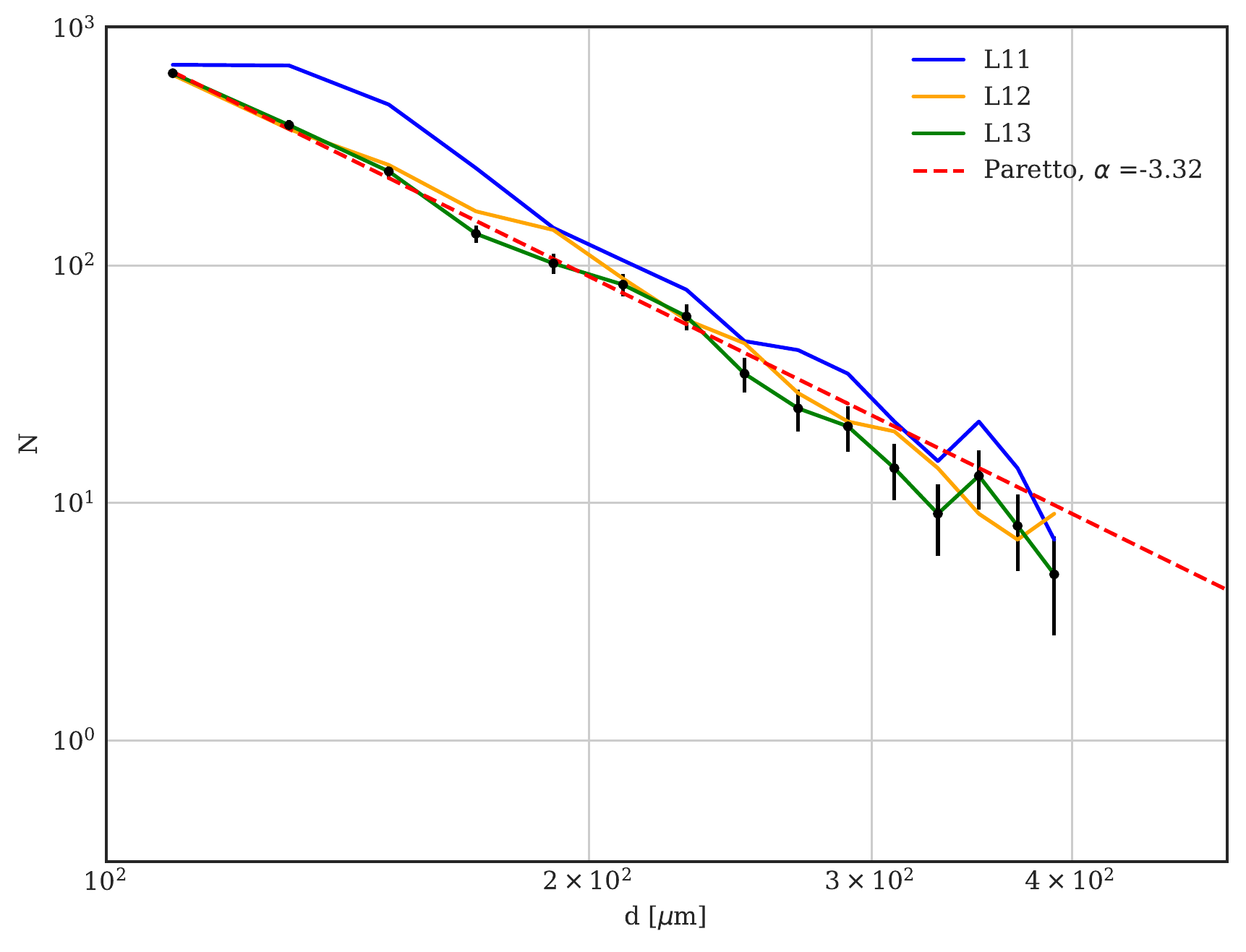}
\end{center}
\caption{Droplet counts in each bin, proportional to the Probability Distribution Function of droplet diameters, for $t=9$ ms and for two grid resolutions L13: $\Delta x=18 \mu $m,  L12: $\Delta x=36 \mu $m and L11: $\Delta x=72 \mu $m. 
\label{fig:pdf}}
\end{figure}
Because of the presence of some non-spherical droplets in the sample, we compared the filtered counts, incorporating only near-spherical droplets using a sphericity index. We find that this filtering removes droplets with $d > 2$mm. (There are less than ten such droplets.)
The resulting number frequency is shown
in Fig. \ref{fig:pdf} together with the statistical error bars, the error being
defined as one standard deviation of the binomial. It is seen that the distribution is close to the power law
$N(d) \sim d^{-3.32}$ for small sizes and then inches down. On Fig. \ref{fig:pdf} we do not show droplet counts for $d < 100$ microns and $d>400$ microns, since the small sizes are plagued by grid resolution errors and the larger sizes by statistical errors.

\begin{figure}[!h]
\begin{center}
  \includegraphics[width=0.4\textwidth]{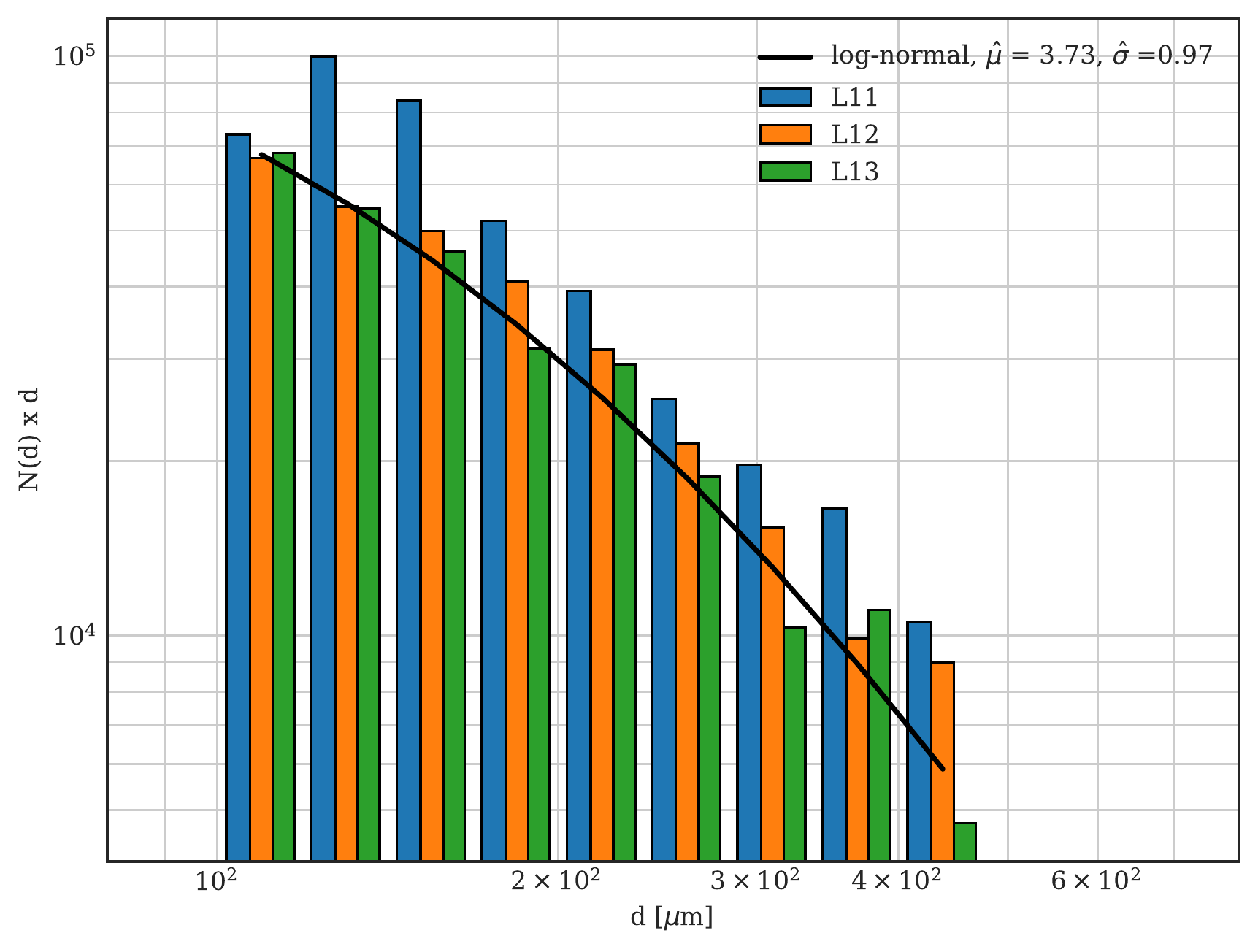}
\end{center}
\caption{Droplet counts in ``log-normal coordinates'' in which the log-normal NF appears as a parabola, with labels L11-13 as in Fig. \ref{fig:pdf}.
\label{fig:pdf2}}
\end{figure}

An often considered number frequency (NF) model is the log-normal, which reads
$ N(d) = ({B}/{d}) \exp\left[ - { (\ln d - \hat \mu)^2}/{(2 \hat \sigma^2)} \right]$  where  $B$ is a normalization constant, $\hat \mu$ is the expected value of $\ln d$, also called the {\it geometric mean}, and
$\sigma$ is the standard deviation of  $\ln d$, also called the {\it geometric standard deviation} (GSD, see \cite{Nicas:2017eo}). If we plot $y= d P(d)$ versus $x = \ln d$ the Log-Normal frequency distribution appears as a parabola. This is done in Fig. \ref{fig:pdf2}. 
That figure shows a fit with a GSD $\hat \sigma = 0.98$. The dimensionless GSD is of the same order of magnitude as the GSD  ($\hat \sigma  \sim 1$) in similar experiments and simulations \cite{Herrmann:2011fq,ling17,Marty:2015vs}, but in the authors' opinion, this may not capture universal underlying physics but the similar range of scales that is within numerical or experimental reach in the literature cited. 
Comparing the NF at various resolution in \cite{Herrmann:2011fq,ling17} it is seen that as the grid size is
reduced the NF shifts to the left, with its geometric mean decreasing and its GSD increasing. This also seen partially in Fig. \ref{fig:pdf} where the L11 resolution peaks ``before'' the two others. It is thus possible that as resolution is increased the curved NF seen on Figs \ref{fig:pdf} and \ref{fig:pdf2} would progressively asymptote to a power law, that is a Pareto, instead of a converged Log-Normal. 
Such a Pareto NF necessarily has  a lower bound, if only the molecular size. This lower bound is not attainable numerically or perhaps even experimentally. A tempting hypothesis is to associate it to the thickness of the sheets as they break or perforate, which involves mechanisms that are not modelled in this study.

To conclude, we have demonstrated a simple physical analog of the physiological mechanism of coughing and sneezing, that is strongly reminiscent of planar sheet atomization processes. A Pareto $d^{-2}$ distribution was not found  but is not excluded at very small diameters. Perspectives include higher resolutions simulations and laboratory experiments in the regime of this numerical experiment, and using the characteristics of the numerically estimated droplet sizes and velocity to predict the further evolution of the droplet cloud using Lagrangian particle methods such as those of Chong et al. \cite{chong2020extended}.

We acknowledge the funding of the Bec.ar programme, the ERC-ADV grant TRUFLOW, the ANR NANODROP grant funded by the {\em Fondation de France} and the PRACE Covid grant of computer time. We thank Lydia Bourouiba, Pallav Kant, Detlef Lohse and St\'ephane Popinet for stimulating and fruitful discussions, and for sharing useful material, visualisations and codes. 
\bibliography{multiphase.bib,biblio.bib}

\end{document}